\documentclass[twocolumn]{aastex631}

\usepackage{xcolor}         
\usepackage{amsmath}
\usepackage{amssymb}
\usepackage{booktabs}       
\usepackage{amsfonts}       
\usepackage{xspace}
\usepackage{mathtools}

\newcommand{\spender}{\textsc{spender}\xspace}

\begin{document}

\title{Autoencoding Galaxy Spectra II: Redshift Invariance and Outlier Detection}

\correspondingauthor{Yan Liang}
\email{yanliang@princeton.edu}

\author[0000-0002-1001-1235]{Yan Liang}
\affiliation{Department of Astrophysical Sciences, Princeton University, Princeton, NJ 08544, USA}

\author[0000-0001-9592-4190]{Peter Melchior}
\affiliation{Department of Astrophysical Sciences, Princeton University, Princeton, NJ 08544, USA}
\affiliation{Center for Statistics \& Machine Learning, Princeton University, Princeton, NJ 08544, USA}

\author[0000-0002-8814-1670]{Sicong Lu}
\affiliation{Department of Physics and Astronomy, University of Pennsylvania, Philadelphia, PA 19104, USA}

\author[0000-0003-4700-663X]{Andy Goulding}
\affiliation{Department of Astrophysical Sciences, Princeton University, Princeton, NJ 08544, USA}

\author[0000-0002-4557-6682]{Charlotte Ward}
\affiliation{Department of Astrophysical Sciences, Princeton University, Princeton, NJ 08544, USA}



\begin{abstract}
  We present an unsupervised outlier detection method for galaxy spectra based on the spectrum autoencoder architecture \spender, which reliably captures spectral features and provides highly realistic reconstructions for SDSS galaxy spectra. We interpret the sample density in the autoencoder latent space as a probability distribution, and identify outliers as low-probability objects with a normalizing flow. However, we found that the latent-space position is not, as expected from the architecture, redshift invariant, which introduces stochasticity into the latent space and the outlier detection method. We solve this problem by adding two novel loss terms during training, which explicitly link latent-space distances to data-space distances, preserving locality in the autoencoding process. Minimizing the additional losses leads to a redshift-invariant, non-degenerate latent space distribution with clear separations between common and anomalous data.
  We inspect the spectra with the lowest probability and find them to include blends with foreground stars, extremely reddened galaxies, galaxy pairs and triples, and stars that are misclassified as galaxies. We  release the newly trained \spender model and the latent-space probability for the entire SDSS-I galaxy sample to aid further investigations.


\end{abstract}



\section{Introduction}

Spectroscopy is a critical tool to probe the physical mechanisms that drive the formation and evolution of galaxies. 
Despite the vast amount of available galaxy spectra provided by large spectroscopic surveys, extracting physical knowledge from them is still a difficult task.
Ideally, one would infer galaxy properties by directly fitting the observed spectrum to a theoretical model, but analytical models are not yet sophisticated enough to reproduce typical individual high S/N galaxy spectra, especially the strong emission lines \citep{tojeiro2007recovering,tojeiro2011stellar}. The physical processes contributing to the observed spectral features may be still poorly understood, thus using oversimplified models could lead to a biased interpretation of the data. 

Alternatively, one may construct a fully data-driven model via unsupervised learning. The main challenge in this approach is to properly disentangle the intrinsic physical spectra from redshift (causing a stretching of the spectra), noise level, and artifacts such as telluric contamination.
Linear models, i.e. a combination of empirical or theoretical templates \citep{brown2014atlas,polletta2007spectral},  are commonly used for redshift estimation and spectral classification \citep{ross2020completed,pace2019resolved,Bolton2012AJ}. The reconstruction  power of linear models is limited by template quality, and often requires many components to achieve a good fit. On the other hand, machine learning (ML) techniques, such as unsupervised random forest \citep{baron2017weirdest} and self-organizing maps \citep{hoyle2015anomaly}, are used for outlier detection without producing a reconstruction.

 \begin{figure*}[ht!] 
  \includegraphics[trim={0 0 0 1cm},clip,width=1.0\linewidth]{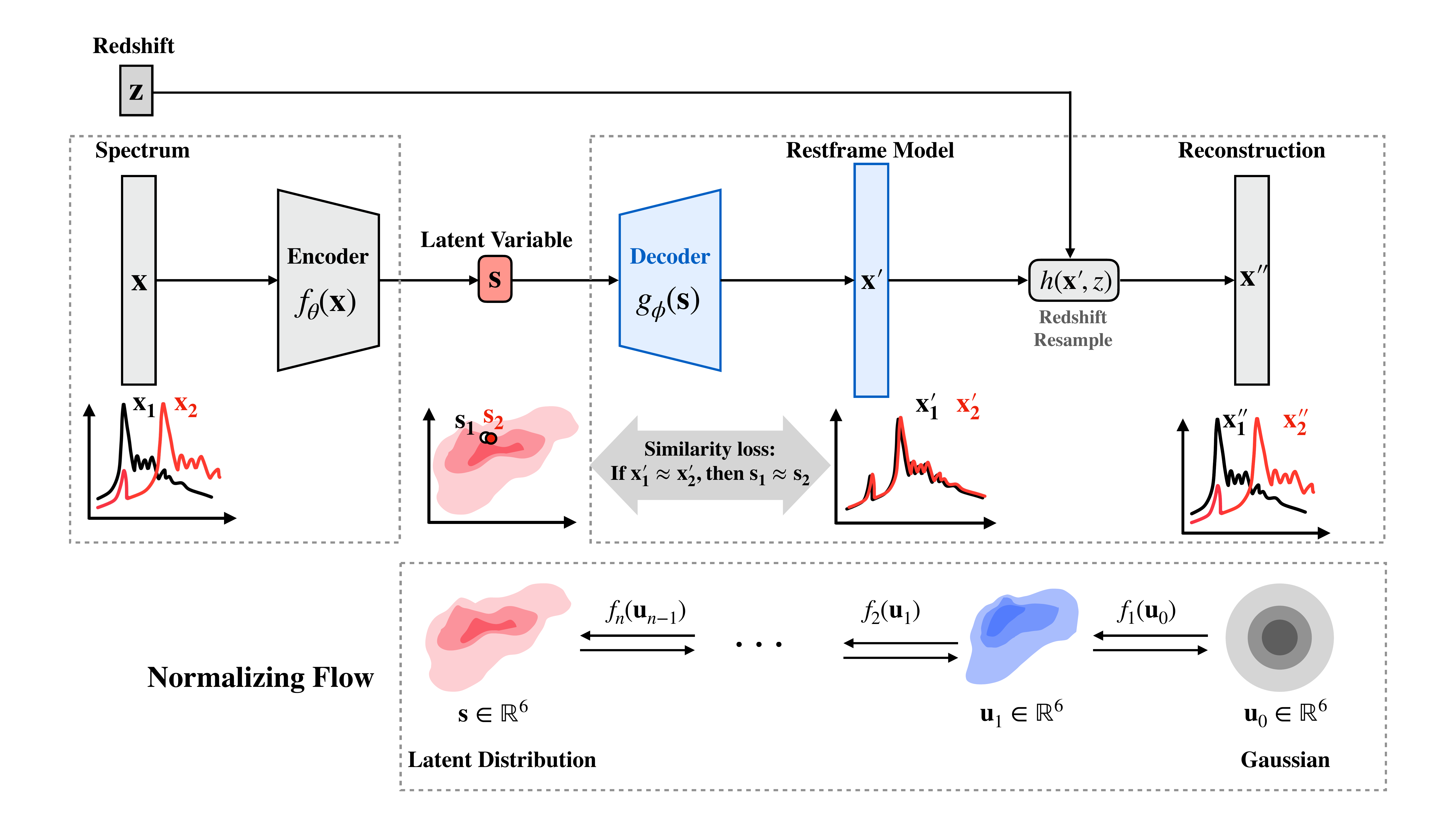}
  \centering
  \caption{The autoencoder \spender generates a restframe and observed spectrum reconstruction. It is trained with a conventional fidelity loss, which compares separately input spectra $(\mathbf{x}_1,\mathbf{x}_2)$ to their redshifted and resampled reconstructions $(\mathbf{x}_1'',\mathbf{x}_2'')$. In this work, we added a novel similarity loss, which links the distance in latent space $|\mathbf{s}_1-\mathbf{s}_2|$ between two spectra to their distance in restframe $|\mathbf{x}_1'-\mathbf{x}_2'|$.
  Spectra of two physically similar galaxies observed at different redshifts $(z_1,z_2)$, for which the underlying restframe models are very similar, will thus yield similar latent vectors.
  To further improve the encoding stability, we minimize an additional consistency loss defined as the latent distances between the input spectra and its artificially red-shifted augment $|\mathbf{s}_{\rm i}-\mathbf{s}_{\rm i,aug}|$. 
  After training the autoencoder,
  we train a normalizing flow to map the complicated latent distribution of the entire SDSS galaxy sample to a Gaussian base distribution through a series of invertible transformations, which allows efficient computation of the probability of any given sample.
  }\label{fig:diagram}
\end{figure*}

Autoencoders (AE) can yield models with good fidelity and small latent dimensionality \citep{portillo2020dimensionality}. 
But for conventional AEs (and other unsupervised ML methods), all galaxy spectra needed to be de-redshifted and resampled to a common restframe, restricting either redshift or wavelength ranges that can be probed.
In an earlier work, we showed that by explicitly adding a redshift transformation to the decoder path, one can utilize the entire spectrum for galaxies at all redshifts \citep{melchior2022}.
This new architecture called \spender effectively compresses SDSS spectra into a low-dimensional latent space, where positions should be invariant under changes of the redshift because they are meant to encode galaxy ``types", with the redshift transformation only affecting the last stage of the generator (see \autoref{fig:diagram}).
In \autoref{sec:method}, we demonstrate this expectation to be na\"ive.
Because of the flexibility of the generator network, different latent-space positions can map to very similar restframe spectra. 
It enables the autoencoder to adopt multiple solutions for representing the same galaxy spectrum, with no penalty in case it chooses a different one at a different redshift.

We then propose a strategy to lift this degeneracy with training on augmented spectra, whose redshifts we have altered, and a similarity metric that relates latent-space distances to distances between reconstructed spectra. At very minor reduction in fidelity, minimizing the additional losses makes the latent space robustly invariant to changes in redshift, observation windows, and noise levels. The encoded parameters become directly interpretable---the spectra of physically similar galaxies cluster in latent space, and positions in the same neighborhood reconstruct similar-looking spectra. 

The interpretability of the latent space enables the discovery of trends and clusters in high-dimensional data, and the characterization of individual objects in a meaningful context (\autoref{sec:latent-space}).  It further allows for the identification of rare or extreme objects in the sample that could potentially reveal new physics, which we will investigate in \autoref{sec:result:ad}.
We also point out, but do not explore in this work, that our new latent space representation supports the creation of realistic mock spectra through sampling from the latent space.
We conclude in \autoref{sec:conclude} with a discussion of the potential uses of this method and the prospects of extending it to future surveys.

\section{Data} \label{sec:data}
We select $\sim$ 500,000 galaxy spectra spanning redshift $z\in[0,0.5]$ from  Sloan Digital Sky Surveys Data Release 16 \citep{ahumada202016th}. Our sample includes all optical spectra that are classified as galaxies and have redshift error $z_{\rm err}<10^{-4}$. Approximately 70\% of the samples are used for training, 15\% for validation, and 15\% is held for testing.
All spectra are zero-padded to a homogeneous wavelength $\lambda_{\rm obs}=3784 \dots 9332$\AA. We use inverse variance weights and masks provided by SDSS Data Release 16 as flux uncertainties. Additionally, we mask out telluric contamination by assigning zero weights to within 5\AA\ of the top $\sim$100 telluric lines, amounting to 12\% of the data vectors. All spectra are normalized by the median flux over restframe wavelengths $\lambda_{\rm rest}=5300\dots5850$\AA. The restframe normalization is chosen to be relatively quiet and accessible at all redshifts to avoid redshift-dependent amplitudes.

\section{Method}\label{sec:method}

Our model architecture is taken from \cite{melchior2022} (see \autoref{fig:diagram} for an overview and the aforementioned paper for details).
Let $\mathbf{x}\in \mathbb{R}^{M}$ denote an input spectrum with $M=3921$ elements. 
It gets encoded, by a modified version of the CNN encoder from \cite{serra2018towards}, into a low-dimensional latent representation, $\mathbf{s}\in\mathbb{R}^S$.
While the architecture can have arbitrary latent dimensions, we choose $S=6$ which provides a good trade-off between the reconstruction quality and model complexity, as suggested by \cite{melchior2022}.
A standard MLP with (256, 512, 1024) nodes and a leaky ReLU activation generates a restframe spectrum $\mathbf{x'}$, whose 7000 spectral elements are chosen to create a mildly super-resolved representation with an extended wavelength coverage ($\lambda_{\rm rest}= 2359\,...\,9332$\,\AA). The reconstructed spectrum $\mathbf{x''}$ is then redshifted and linearly interpolated from $\mathbf{x'}$ to the same wavelengths as $\mathbf{x}$.

\begin{figure} [t]
  \includegraphics[width=\linewidth]{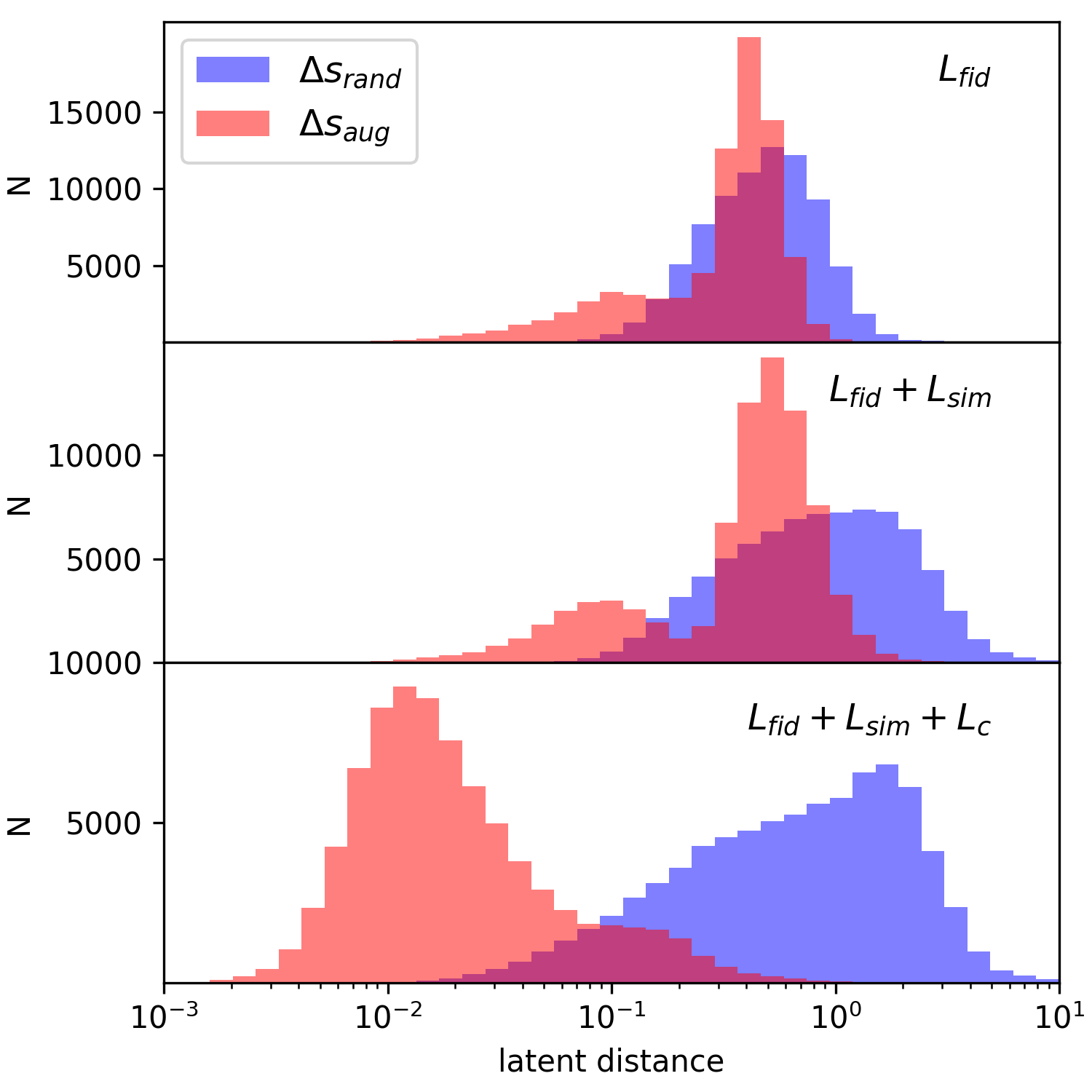}
  \centering
  \caption{Euclidean distance distribution in latent space for randomly selected galaxy pairs ($\Delta \mathbf{s}_\mathrm{rand}$, blue) and original-augment pairs ($\Delta \mathbf{s}_\mathrm{aug}$). Augmented spectra stem from the original spectra by artificially altering the redshifts. If the encoder is redshift invariant, $\Delta \mathbf{s}_\mathrm{aug}=0$.
  The three panels refer to the result of training with different terms in the loss function \autoref{eq:loss}. Optimizing the full loss function (bottom)  significantly reduces $\Delta \mathbf{s}_\mathrm{aug}$, while increasing $\Delta\mathbf{s}_{\rm rand}$, i.e. increasing the separation of different spectral types in latent space.
  }\label{fig:s-distance}
\end{figure}

This architecture allows for a redshift-invariant latent encoding because the redshift transformation is explicitly performed in the generator. 
One might therefore expect that the network will use the same latent variables to represent the same restframe spectrum at all redshifts as that would simplify the decoding process. However, we find that the encoder network does not typically yield the same latent in this case. 
We arrive at this unintuitive finding by encoding ``augmented" galaxy spectra, for which we have artificially altered the redshifts.
The top panel of \autoref{fig:s-distance} shows the distribution of Euclidean distances in latent space $\Delta \mathbf{s}_\mathrm{aug}$ between the original and the augmented spectrum (in red), which has a comparable spread to the distance distribution of randomly selected original spectra (in blue).
We interpret this result as the high flexibility of the architecture allowing the decoder to reconstruct indistinguishable models from different encoded locations. 
To achieve the best reconstructions across a range of spectral types and redshifts, a redshift-dependent encoding seems preferable.

\begin{table*}[t]
  \centering
  \begin{tabular}{lcccccc}
    \toprule
    Training objective & $L_{\rm fid}$ & $L_{\rm sim}$& $L_c$ &  $\langle \Delta \mathbf{s}_\mathrm{aug}\rangle$ &
    $\langle \Delta \mathbf{s}_\mathrm{rand}\rangle$ &
    $\langle \Delta \mathbf{s}_\mathrm{aug}\rangle / \langle \Delta \mathbf{s}_\mathrm{rand}\rangle$\\
    \midrule
    $L_{\rm fid}$ & 0.387 & - & -  & 0.353 & 0.540 & 65.4\%\\
    $L_{\rm fid}+L_{\rm sim}$  &  0.388 & 0.177  & - &0.483 & 1.232 &39.2\%\\
    $L_{\rm fid}+L_{\rm sim}+L_c$ & 0.389 & 0.177 & 0.020 & 0.043 & 1.103 &3.9\%\\
    \bottomrule
    \\
  \end{tabular}
    \caption{Training performance using different strategies, evaluated on the test set. $\langle \Delta \mathbf{s}_\mathrm{aug}\rangle$ denotes the average Euclidean distance between encoded data-augment pairs, $\langle \Delta \mathbf{s}_\mathrm{rand}\rangle$ for randomly chosen data pairs. See \autoref{fig:s-distance} for a visualization of the same data. $L_{\rm sim}$ is evaluated assuming $k_0=2.5, k_1=5.0$. Note that when the latent distance and spectral distance are perfectly aligned, the minimum similarity loss $L_{\rm sim}$ is 0.151.}
  \label{tab:performance}
\end{table*}

\subsection{Loss Function}

Our basic training in \citet{melchior2022} used a conventional fidelity loss, which quantifies the reconstruction quality, assuming normally distributed noise. It measures the mean log-likelihood of the reconstruction of spectral elements averaged over batches of $N$ spectra with spectral size $M$:
\begin{align}
L_{\rm fid} &= \frac{1}{2NM}  \sum_i^{N}  \mathbf{w}_{i}\odot(\mathbf{x}_i-\mathbf{x}_i'')^2,
\end{align}
where $\mathbf{x}_i $ is the $i$-th input spectrum, $\mathbf{w}_i$ its inverse variance, and $\odot$ the element-wise multiplication. 

\begin{figure*} [t]
  \includegraphics[width=\linewidth]{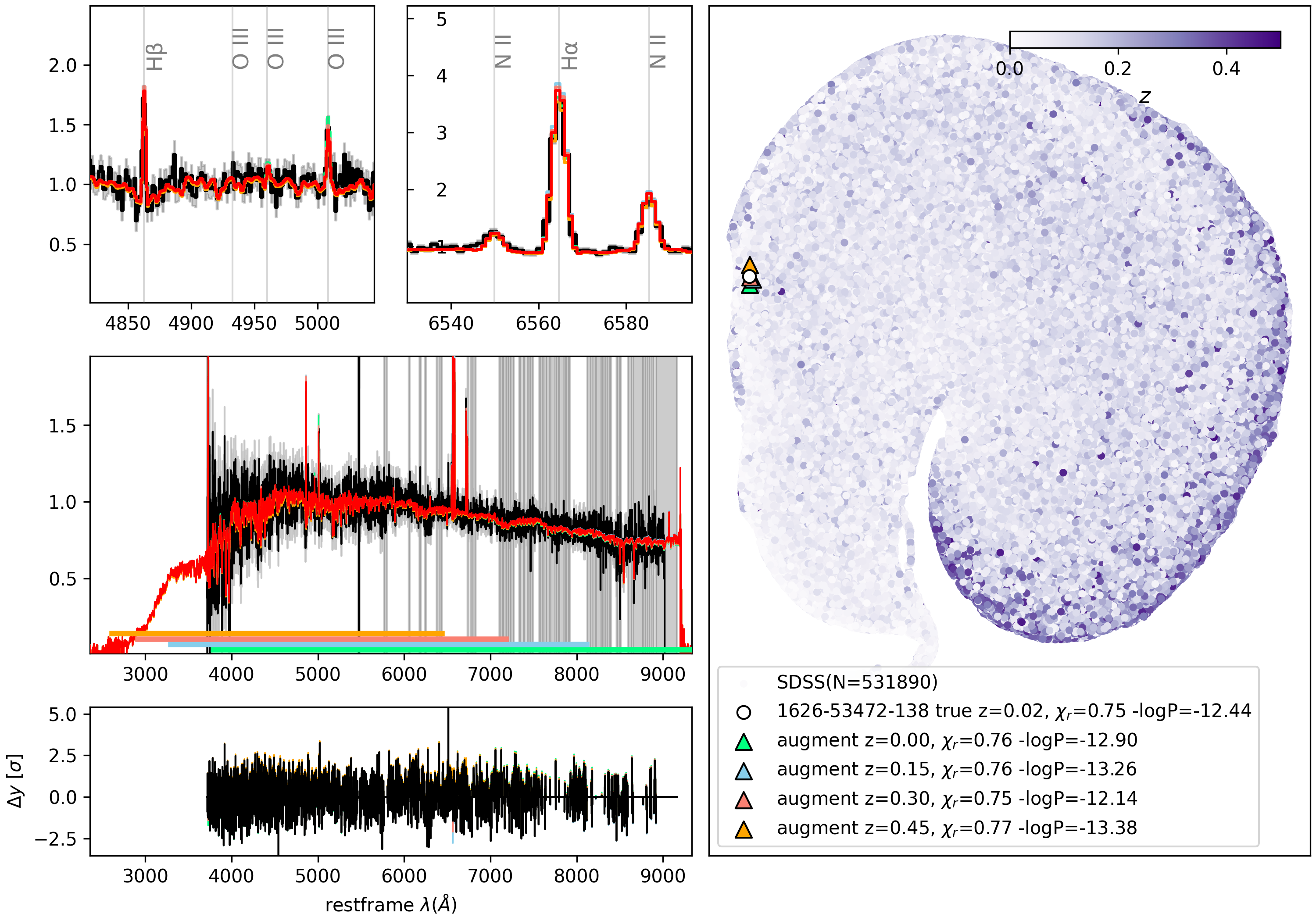}
  \centering
  \caption{\emph{Left:} Observed spectrum (black) of a galaxy from the test sample at $z=0.02$, its reconstruction (red), and reconstructions of augmented spectra with artificially altered redshifts (color-coded). Zoomed-in versions are shown on the top. The colored bar in the middle-left shows the observed wavelength range of the augments. \emph{Right:} UMAP-embedded latent space distribution of the entire SDSS galaxy sample. The white circle marks the example spectrum, and the triangles mark its augments.
  }\label{fig:spectrum}
\end{figure*}

A stronger constraint is required to generate robustly redshift-invariant encoding.
Our extended loss function
\begin{equation}
\label{eq:loss}
L_{\rm total}=L_{\rm fid} + L_{\rm sim} + L_{\rm c}
\end{equation}
operates on all four of the stages shown in \autoref{fig:diagram}.
We define a similarity loss $L_{\rm sim}$ that, unlike the fidelity loss, operates on the two intermediate stages. Let $\mathbf{s} = f_\theta(\mathbf{x})$ be
the encoded latent vector and $\mathbf{x'} = g_\phi(\mathbf{s})$ be the restframe model, where $f_\theta, g_\phi$ are parameterized encoder and decoder functions. Ideally, if two restframe models, $\mathbf{x}_i'$ and $\mathbf{x}_j'$ are similar, their latent positions $\mathbf{s}_i$ and $\mathbf{s}_j$ should be similar as well; otherwise, the mapping from latent space to spectrum space becomes difficult to interpret. On the other hand, distinctively different models should have well-separated latent positions. 
The desired relation can be satisfied naturally if the latent distance is proportional to a spectral distance: $|\mathbf{s}_1-\mathbf{s}_2|^2 \propto |\mathbf{x}_1'-\mathbf{x}_2'|^2$, which motivates us to set up a loss term as follows:
\begin{equation}
\label{eq:Lsim}
\begin{split}
L_{\rm sim}(\mathbf{S}, k_0, k_1) = 
\frac{1}{N^2}\sum_{i,j}^N  &{\rm sigmoid}(k_1  \mathbf{S}_{i,j}-k_0) \\
+ \frac{1}{N^2}\sum_{i,j}^N  &{\rm sigmoid}(-k_1 \mathbf{S}_{i,j}-k_0),
\end{split}
\end{equation}
where
\begin{equation}
\label{eq:Sij}
\mathbf{S}_{i,j} =\frac{1}{S}|\mathbf{s}_i-\mathbf{s}_j|^2
- \frac{1}{M} |\mathbf{w}'\odot(\mathbf{x}_i'-\mathbf{x}_j')|^2,   
\end{equation}
$\mathbf{w}'$ is the inverse-variance weight defined at restframe wavelengths, and $k_0,k_1$ are adjustable hyper-parameters that control the steepness of the slope. 
$L_{\rm sim}$ encourages pairwise latent space distances proportional to the spectra (dis)similarity defined in \autoref{eq:Sij}.
The double sigmoid function in \autoref{eq:Lsim} serves two purposes. It limits the effect of the similarity loss by setting $L_{\rm sim} \leq 1$, such that it is only important when the fidelity loss is also comparably low, $L_{\rm fid} \lesssim 1$. And it provides relatively smooth gradients (compared to e.g. $\mathbf{S}_{i,j}^2$ ), improving the trainability of the model.
It is crucial to measure the similarity between restframe pairs rather than input pairs, because the former provides a stable measure independent of redshift and observational window. In addition, restframe models usually contain less noise than the original data (see \autoref{fig:spectrum}, bottom left panel).

Inspired by \citep{sinha2021consistency}, we add the consistency loss as a more guided form of the similarity loss: 
\begin{align}
L_{\rm c} &= \frac{1}{N}\sum_i^N {\rm sigmoid}\left[\frac{1}{\sigma_s^2 S}|\mathbf{s}_i-\mathbf{s}_{\mathrm{aug},i}|^2\right]-0.5
\end{align}
where $\mathbf{s}_i$ and $\mathbf{s}_{\mathrm{aug},i}$ are the latent positions of the original and its corresponding augment spectrum, that has been redshifted by a random $z_{\rm new}$ from a uniform distribution between $[0,0.5]$. In practice, we set $\sigma_s =0.1$ which provides a good trade-off between the consistency and the overall scale of the latent distribution. Minimizing $L_\mathrm{c}$ reduces the latent distances between galaxy spectra and their augments, improving the encoding stability against redshift variation. The consistency loss reinforces that physically similar spectra pairs are pulled together in latent space, while dis-similar pairs are moved apart by the (dis)-similarity loss term \autoref{eq:Sij}.
This combination of losses should make the latent space also more expressive by encouraging of clustering of similar and separation of dissimilar spectra, which we will exploit in \autoref{sec:result:ad}.


\subsection{Training}
\label{sec:training}

To evaluate the impact of different loss terms, we trained three models using different strategies: In the first training, we optimize the model with $L_{\rm fid}$ alone to serve as the baseline. In the second model, we optimize $L_{\rm fid}+L_{\rm sim}$ simultaneously, with an ``inverse-annealing'' cycle to gradually increase the slope $k_1$, while $k_0$ is held constant. For the third model, we add the consistency regularization and optimize the full loss defined in \autoref{eq:loss}.

We train each model on an NVIDIA A100 GPU using the same training and validation data set for 1000 epochs and observe convergence. 
The results are summarized in \autoref{tab:performance} and \autoref{fig:s-distance}. With the fidelity loss being essentially identical in all three runs, the best performance in terms of redshift invariance is achieved with all three losses combined. 
In particular, including the similarity loss $L_{\rm sim}$ encourages dis-similar spectra pairs to move apart (see \autoref{fig:s-distance}, middle panel), increasing the average latent distance between random spectra pairs by a factor of 2. 
Optimizing $L_{\rm c}$ along with $L_{\rm fid}+L_{\rm sim}$ reduces the average latent distance of augmented redshifted spectra by a factor of $\sim$10 (\autoref{fig:s-distance}, bottom panel). As a result, in the latent space of the best-fit model, the true and augmented spectra are 25 times closer to each other than the average pair-wise distance, while for the fidelity-alone model, the original and augmented spectra are only 1.5 times closer than the average. This order-of-magnitude improvement has been achieved with a very minor decrease in fidelity, which suggests that the autoencoder indeed has sufficient flexibility to accommodate the additional demands to the latent space without sacrificing reconstruction quality.

We suggest that the consistency loss should not be used without the similarity loss because it merely seeks to compress the volume occupied by the latent distribution and collapse it to a point at the origin. Instead, it is best added once a model trained with fidelity and similarity losses is available. 

With the best model, we achieve a desirable redshift-invariant behavior. \autoref{fig:spectrum} shows the reconstructed restframe model and encoded latent position for an example galaxy. The latent positions of the original and redshift-augmented spectra are stably grouped together (colored markers in the right panel), even for the $z=0.45$ augment (colored in orange) where the dominant H$\alpha$ emission line is outside of the observable wavelength range, i.e. unavailable to the encoder. The reconstruction quality is evidently robust to such missing features.

\section{Latent Space Structure}
\label{sec:latent-space}

\begin{figure*} [t]
  \includegraphics[width=0.8\linewidth]{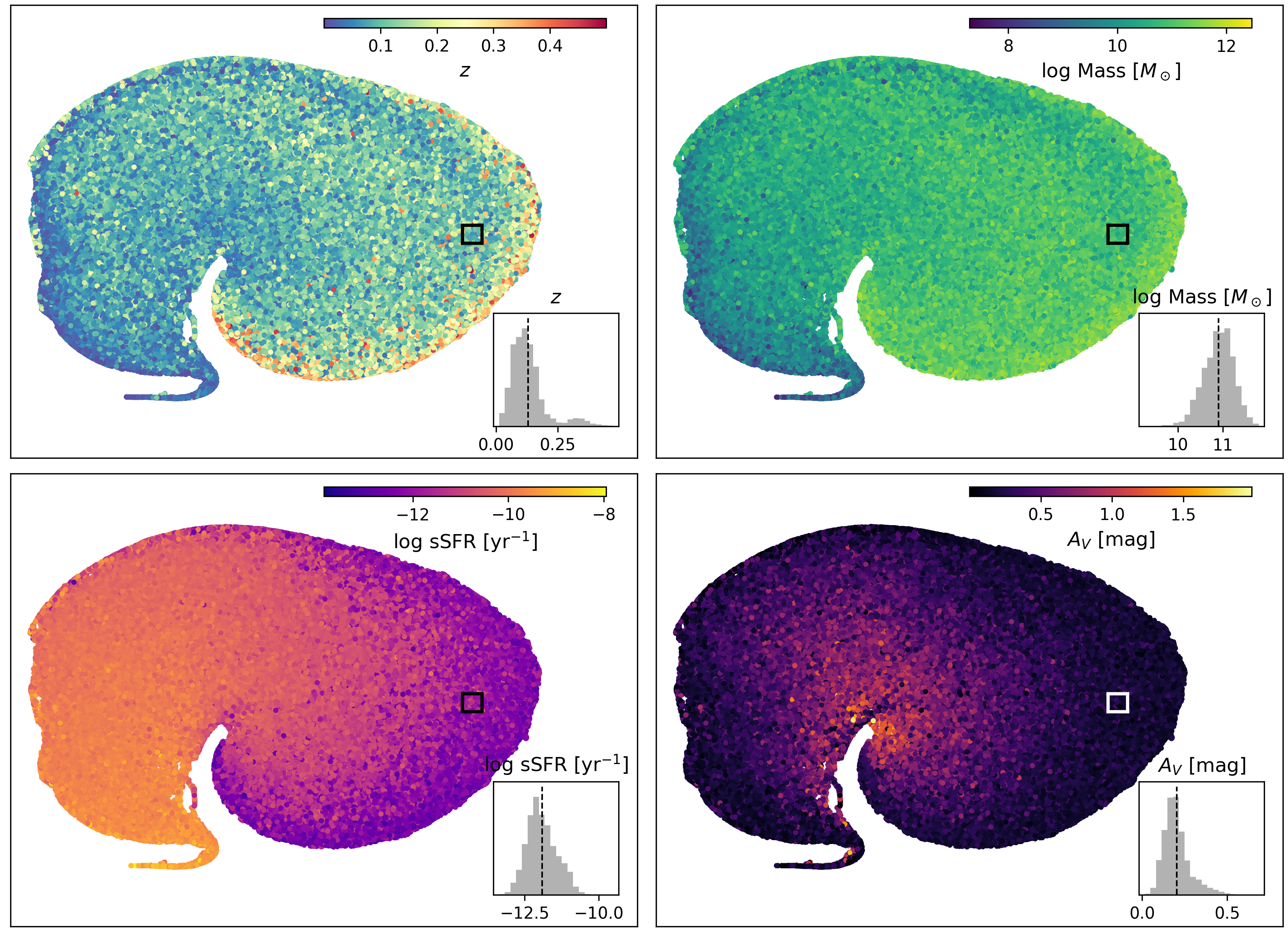}
  \centering
  \caption{UMAP embedding of the latent variables for all 531,890 SDSS galaxy spectra, color-coded by redshifts (top left), stellar masses (top right), specific star formation rates (lower left), and dust attenuation in restframe V band (lower right). The lower right histograms show the distribution of physical properties of $\sim$2000 samples located inside a small box (marked black/white) in the UMAP space.
  }\label{fig:umap}
\end{figure*}


Following the approach we took in \citet{melchior2022}, we inspect the latent space distribution by further projecting the latent variables of the entire galaxy spectra sample into a 2-dimensional UMAP embedding \citep{mcinnes2018umap}.
As shown in \autoref{fig:umap}, the UMAP embedding of the dataset is continuous and connected.
To test our assumption that the latent space encodes the physical properties of the galaxy without our supervision, we plot the stellar masses, star formation rates, and dust attenuation estimates from the GALEX-SDSS-WISE LEGACY CATALOG \citep{Salim2016GALEX,Salim2018Dust}, which covers $\sim 80\%$ of the SDSS galaxy spectra. 

As shown in \autoref{fig:umap}, the 2-dimensional embedding of the latent space strongly correlates with each of the physical properties: the specific star formation rate is predominantly following a continuous trend in the counter-clockwise direction, from the most quiescent galaxies to starburst galaxies (bottom left panel). The stellar mass of galaxies correlates with the clockwise direction (top right panel). The galaxies with the strongest dust attenuation seem to cluster around a ``locus'' where the star-forming and quiescent galaxies are squeezed together, and the attenuation weakens inside-out in the radial direction (bottom right panel). 


One might notice that the stellar mass appears to correlate with redshift (top left panel). As we have argued in \citet{melchior2022}, this is an artifact of a magnitude-limited sample. At high-redshift, only the most massive galaxies are observed, while low-mass galaxies can be observed only at low redshifts.

By inspecting the latent space structures, we conclude that our latent space is well-behaved and physically informative. One can, for instance, assign an estimate of the physical properties of an SDSS-I spectrum according to its neighbors in latent space.
Similar approaches have been reported with direct embeddings, but have so far been restricted to e.g. lower-dimensional spaces of broad-band SEDs \citep{Hemmati2019-fh} or to spectra at $z\approx 0$ \citep{Traven2017-xp}. We see this capability as beneficial for subpopulations of galaxies where conventional spectrum analysis does not produce reliable parameter estimates, or none at all.
Without the extended loss during training introduced in \autoref{sec:method}, the association between latent space position and physical parameters would be subject to random scatter from variations in redshift.

\begin{figure*}[t!]
  \includegraphics[width=0.7\linewidth]{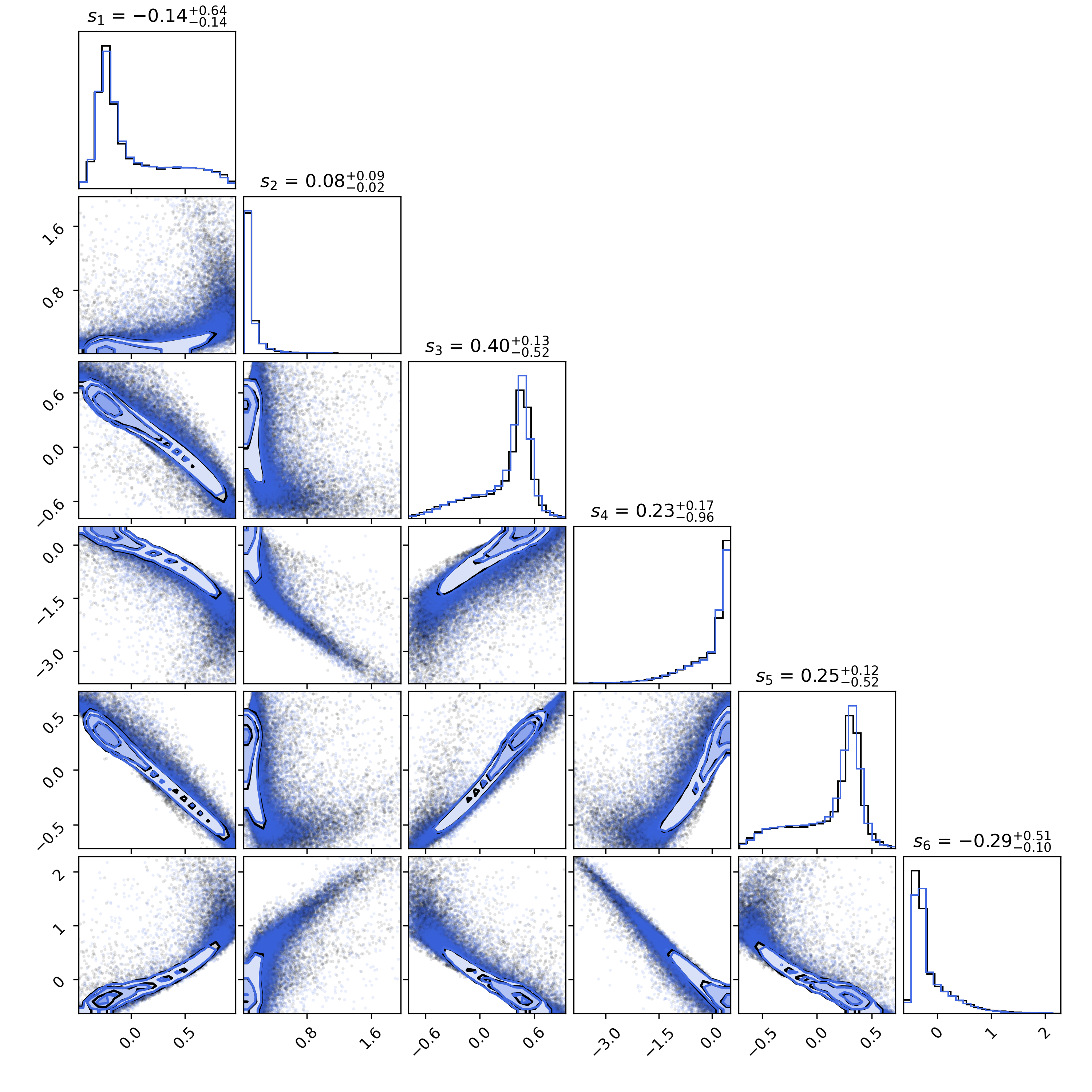}
  \centering
  \caption{The latent space distribution of the validation data (black) and the distribution of the mock samples generated by the normalizing flow (blue) for each of the 6 latent variables.
  }\label{fig:nf}
\end{figure*}

\section{Outlier Detection} \label{sec:result:ad}

Because of our choice of additional loss terms, we surmise that the latent space is also well-suited for outlier detection, as the anomalous samples will be pushed away from more typical ones.
However, despite the apparent simplicity of the UMAP embedding in \autoref{fig:umap}, the latent space has a complex distribution in $\mathbb{R}^S$.
Determining outliers in this space is not trivial.
We therefore implement an additional mapping from the latent distribution of the SDSS galaxy sample to a $S$-dimensional normal distribution using a normalizing flow.

\subsection{Normalizing Flow}
\label{sec:nf}

A normalizing flow \citep[NF,][]{papamakarios2017masked} consists of a series of trainable reversible transformations. 
Let us define $f=f_i \circ f_2 \dots \circ f_n$ as the transformation from the base space $\mathbf{u}$ to target space $\mathbf{s}$. Let $\pi_s(\mathbf{s})$ be the target density and $\pi_u(\mathbf{u})$ be some simple base density (i.e. Gaussian). The resulting model of the probability density in the target space is:
\begin{align}
    p(\mathbf{s})=\pi_u(f^{-1}(\mathbf{s})) \left| \det{\left (\frac{\partial f^{-1}}{\partial \mathbf{s}}\right)} \right |
\end{align}
where $\det{\left (\partial f^{-1}/\partial \mathbf{s}\right)}$ is the Jacobian of the inverse transformation $f^{-1}$. It can be shown that maximizing the likelihood of data $\log{p(\mathbf{s})}$ is equivalent to minimizing the Kullback–Leibler divergence (KL divergence) between the proposed distribution and target distribution $D_{\rm KL}(\pi_s(\mathbf{s})\|p(\mathbf{s}))$.

To implement these transformations, we set up a 5-layer masked auto-regressive flow (MAF) with 50 hidden features using \texttt{nflows} \citep{nflows-Durkan2020} and adopt the total negative log-likelihood as the loss function. 
\begin{align}
 L_{\rm MAF} = \sum_i \log{p(\mathbf{s_i})}\label{eq:nf}
\end{align}
The entire SDSS galaxy samples are encoded by the best pre-trained auto-encoder, with 85\% samples
used for training and 15 \% for validation. 
We train the normalizing flow on an Intel CPU node with 4G RAM using 10,000 samples per batch and observe convergence after 2,000 epochs.

We test the performance of our normalizing flow by taking samples from the base distributions and using the model to transform them into the learned target distribution.
As shown in \autoref{fig:nf}, the distribution of the mock sample generated by the model and the true latent distribution of the validation set (N=104786) are essentially indistinguishable.

\subsection{Top-Ranked Outliers}
\label{sec:outliers}

\begin{figure*}[t!] 
  \includegraphics[trim={0 0 0 1cm},clip,width=0.9\linewidth]{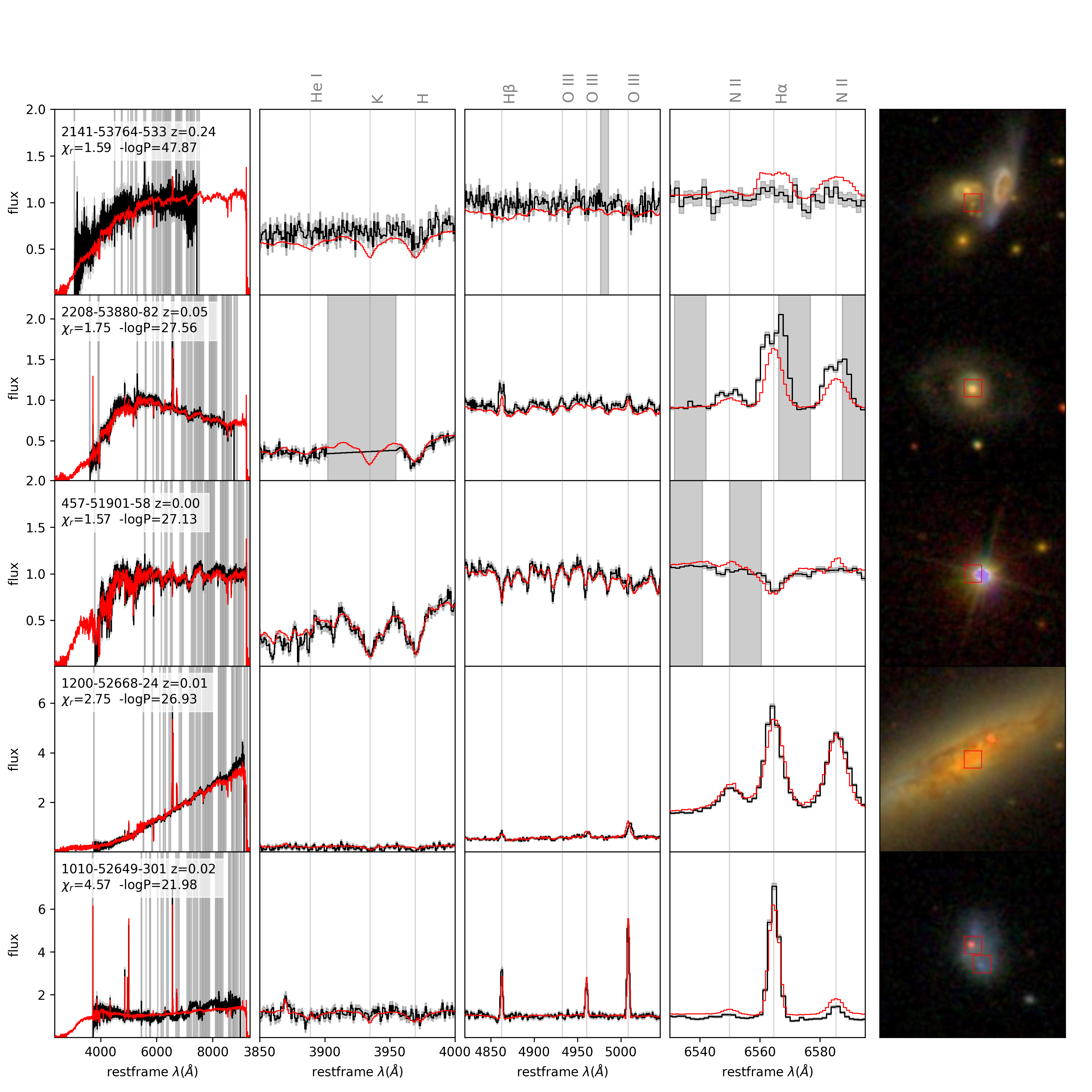}
  \centering
  \caption{A sample of the outliers identified by the normalizing flow, ranked from top to bottom by the assigned probabilities, includes an interacting galaxy system, a double-peaked emission galaxy, a star misclassified as a galaxy, a galaxy with extreme dust attenuation, and a chance alignment of a WD-MS binary and background galaxy.
  The first column shows the entire normalized spectrum (black) with uncertainties (grey shading) and its reconstruction (red). The second, third, and fourth columns zoom in to important emission/absorption lines. The corresponding images are shown in the last column.
  The legend on the top left corner includes the SDSS ID, redshift, reduced chi-square, and the negative log probability of each spectrum.
  }\label{fig:logp}
\end{figure*}

Having confirmed that the normalizing flow accurately reconstructs the encoded parameters of SDSS galaxies, we evaluate the likelihood for each spectrum in the base distribution, where the anomalous data will be assigned low likelihoods. We show a sample of the top-ranked, i.e. lowest probability, outliers in \autoref{fig:logp}.

For further analysis, we focus on the targets with low likelihood but good reconstruction quality, excluding the strongest line emitters, for which the maximum flux is more than 100 times the continuum flux, as the \spender model often fails to accurately reproduce such strong emission lines. 
We visually inspect the SDSS spectra and images of the top 100 objects that are assigned the lowest likelihood, and find a variety of different objects, from blends to supernovae.

The largest group of NF outliers is formed by 44 blended objects---superpositions of multiple galaxies and/or stars. Identifying blends in images can be relatively straightforward, but detecting them from spectroscopic data is often challenging, especially when the spectra lack emission lines. Our method is particularly sensitive to blends, as the encoder could not easily reconcile the line and continuum features of the two objects with its expected spectrum shape of a single object. The NF anomaly detection is able to identify blends as outliers even when the spectra are almost featureless, which suggests that the auto-encoder indeed has a reliable notion of the typical shape of a common continuum. In 31 of the 44 blends, the primary and secondary objects are well-separated in the SDSS image. We present an example of such object, SDSS 2141-53764-533 ($-\log p=47.87$), a complex interacting system with long tidal tails, in the top row of \autoref{fig:logp}. 
In the remaining 13 blending systems, the primary and secondary objects are too close to be distinguished visually. In this case, we first subtract the best \spender model from data, then look for a second set of features in the residual spectrum. We show an example, SDSS 1010-52649-301 ($-\log p=21.98$), with interesting patterns in the residual spectrum in the last row of \autoref{fig:logp}. This system appears in a white dwarf-main sequence binary catalog \citep{liu2012white}, and the strong emission lines suggest that is coincidentally aligned with a background galaxy.

We identified 10 emission-line galaxies with velocity structures, such as double-peaked emissions, and asymmetric lines. When an object with a velocity structure is resolved, we might observe some additional features in the image. For example, SDSS 2208-53880-82 ($-\log p=27.56$) has a ring-like structure that indicates rotational motion (\autoref{fig:logp}). Similarly, another example not shown here, SDSS 617-52072-326 ($-\log p=23.28$), also known as LEDA 2816158, is reported as interacting with a spiral companion (LEDA 2816157), confirmed by the long tidal tail in the image \citep{petrosian2002studies}.

We found two stars that are misclassified as galaxies by the SDSS data pipeline, one of which is shown in the third row of \autoref{fig:logp}. K stars can mimic elliptical galaxies that lack emission lines. It is evident from the image that SDSS 457-51901-58 ($-\log p=27.13$) is a star because it has a morphology consistent with a single point spread function (PSF), including visible diffraction spikes.

We found five extremely dusty objects with almost no blue-side emission. We show one example of a strongly reddened galaxy, SDSS 1200-52668-24 ($-\log p=26.93$),  in the fourth row of \autoref{fig:logp}. Evidently, the H$\beta$ emission is heavily suppressed compared to $H\alpha$, and the corresponding image clearly reveals the presence of dust clouds.

Two outliers, SDSS 2430-53815-267 ($-\log p=18.42$) and SDSS 1581-53149-470 ($-\log p=19.03$), are galaxies hosting supernovae that were active at the time when the spectra were taken. Both of them are reported as type Ia supernovae by \cite{graur2013discovery}. 

Some of the NF outlier spectra were already reported in the literature including:
SDSS 277-51908-519 ($-\log p=18.45$) is identified as a BL Lacertae object by \cite{pena2021optical}. 
SDSS 2745-54231-4 ($-\log p=18.17 $) is selected by \cite{lehmer2022elevated} as a galaxy with unusually high specific star formation rate ($ 3.77_{-0.10}^{+0.14}$ yr$^{-1}$) and low metallicity ($ \sim 0.3 Z_\odot$).
SDSS 2145-54212-388 ($-\log p=19.86$) and SDSS 1998-53433-304 ($-\log p=21.18$) are star-forming galaxies showing strong nebular emission {He\,{\footnotesize II}$\lambda$4686 that is believed to be dominated by emission from Wolf–Rayet stars \citep{shirazi2012strongly}.

We classify five outlier spectra as ``Bad Spectra'' because a significant portion of them are missing, while the remaining parts are simply connected by straight lines.

Lastly, we found five outliers for which we could not find an explanation through visual inspection. However, we note that all of them stem from the same plate, 2208, taken on the same date, MJD 53880. One possible explanation is a subtle calibration mismatch for this plate.

The types of outliers we found are generally consistent with those identified by \cite{baron2017weirdest}. We note that, in comparison, our method appears to be more sensitive to blends as outliers. A possible explanation could be our simultaneous use of line and continuum features,  while their unsupervised random forest method considers each spectral bin separately.

We summarize our findings from the Top-100 outliers with our, admittedly subjective, classification in \autoref{fig:pie}.
Note that there is no motivation for limiting the sample to 100 other than the time it takes to individually inspect and classify each spectrum. The labels we created for the above NF outliers describe only our best guess for its assigned low probability instead of a rigorous and quantitative analysis.
We provide the full Top-100 outlier list in \autoref{tab:top100} and the NF probabilities for the entire SDSS galaxy sample at our code repository, and encourage interested readers to carry out their own follow-up studies.

\begin{figure} [t]
  \includegraphics[width=\linewidth]{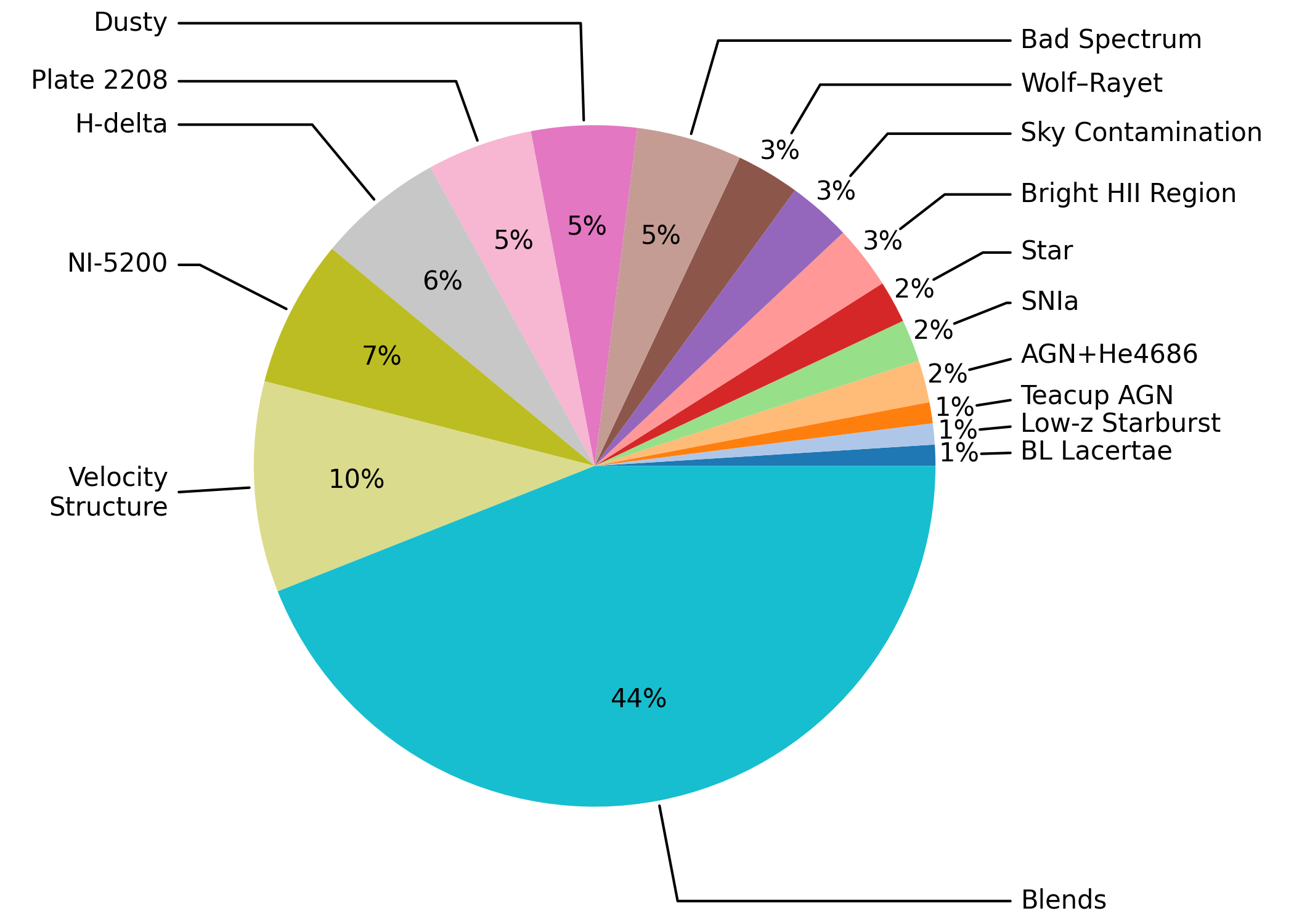}
  \centering
  \caption{Visual classification of 100 SDSS objects that are deemed most unlikely given their position in the \spender latent space.}\label{fig:pie}
\end{figure}

\section{Discussion and Conclusion}
\label{sec:conclude}

In this study, we performed unsupervised outlier detection on an interpretable, redshift-invariant encoding of SDSS galaxy spectra using the \spender architecture. 
Our best auto-encoder model generates highly realistic spectra for the entirety of the SDSS spectroscopic galaxy sample with six latent parameters, which encode the physical properties probed by the galaxy spectra.

We defined a new loss term to directly relate distances in latent space to distances in restframe space, which discourages latent-space degeneracies that act as a source of stochasticity on any subsequent use of the latents. In particular, the decoder can learn a more robust and complete underlying restframe model over the entire redshift range.
In addition, the association of latent-space position to physical properties is more stable and evident with such a robust encoding.

Training with similarity loss encourages physically-similar galaxies to cluster in latent space, as long as there are enough intermediate objects to link them.
This new loss term encouraged an encoding that is well-suited for outlier detection: outliers are pushed away from more common samples, who themselves tended to cluster.

We trained a normalizing flow on the latent distribution of the entire SDSS galaxy sample and identified the top 100 objects that are deemed most unlikely by the normalizing flow model. By visual inspection and classification, we confirmed that our anomaly detection method finds a wide variety of meaningful astrophysical and instrumental outliers.
The most common group of outliers we found were due to the blending of multiple sources, either stars or galaxies.
These blends were present in the SDSS main galaxy sample with high-quality redshifts.
Were they to be used, e.g., in the training of photometric redshift estimators, they could cause erratic and suboptimal performance.

We caveat that outliers are, by definition, at locations in latent space that are sparsely populated. Because very little training data is available, both the encoding and the modeling by the normalizing flow are much less reliable for such objects. We therefore suggest considering the probability $p(\mathbf{s})$ reported by the normalizing flow rather qualitatively: very low values indicate rare objects, but the exact values themselves may not be meaningful.


On the other hand, our immediate reason to utilize a normalizing flow, namely its fast evaluation of the sample likelihood, enables applications beyond outlier detection.
In particular, it allows for rapid sampling from the latent distribution, which, coupled to the decoder, yields mock samples of realistic restframe spectra. As our architecture can adjust redshift, spectral resolution, and line spread function, we can produce mock spectra for instruments or surveys other than SDSS.
Furthermore, because of the clear association of physical parameters with latent-space position, one can estimate those parameters by encoding an unseen spectrum and then adopting the parameters of interest from neighbors in latent space, without the need for a full-fledged spectrum analysis pipeline.
To support such applications, we make our new \spender model as well as the full catalog of latent-space probabilities available at \url{https://github.com/pmelchior/spender}.

\section*{Acknowledgments}
\begin{acknowledgments}

The authors wish to thank Michael Strauss for his swift help with the visual inspection and classification of the outliers discussed in this paper.
This work was supported by the AI Accelerator program of the Schmidt Futures Foundation.
The authors are pleased to acknowledge that the work reported on in this paper was substantially performed using the Princeton Research Computing resources at Princeton University which is a consortium of groups led by the Princeton Institute for Computational Science and Engineering (PICSciE) and Office of Information Technology's Research Computing.

\vspace{1em}

\software{
    \href{https://pytorch.org/}{\texttt{Pytorch}} \citep{paszke2019pytorch},
    \href{https://corner.readthedocs.io/}
    {\texttt{corner}} \citep{corner},
    \href{https://umap-learn.readthedocs.io/}
    {\texttt{umap}} \citep{sainburg2021parametric}
    \href{https://github.com/huggingface/accelerate}{\texttt{accelerate}} \citep{accelerate-Gugger2022},
    \href{https://github.com/aliutkus/torchinterp1d}{\texttt{torchinterp1d}},
    \href{https://github.com/bayesiains/nflows}{\texttt{nflows}} \citep{nflows-Durkan2020},
    \href{http://www.numpy.org}{\texttt{NumPy}} \citep{harris2020array},
    \href{https://www.astropy.org/}{\texttt{Astropy}} \citep{Astropy2022ApJ},
    \href{https://matplotlib.org}{\texttt{Matplotlib}} \citep{Hunter:2007}}

\end{acknowledgments}

%






\begin{table*}[t]
  \centering
  \resizebox{\textwidth}{!}{
  \hspace*{-3cm}\begin{tabular}{lcccl lcccl}
    \toprule
 \multicolumn{5}{c}{Outlier 1$\sim$50} & \multicolumn{5}{c}{Outlier 50$\sim$100} \\
 \cmidrule(lr){1-5}\cmidrule(lr){6-10}
SDSS ID  & RA & DEC & $-\log p$ & Label & SDSS ID  & RA & DEC & $-\log p$ &Label \\
 \cmidrule(lr){1-5}\cmidrule(lr){6-10}
634-52164-68   & 310.92025    & -6.20623     & 84.79    & Blends   & 2268-53682-16  & 122.67157    & 13.11158     & 22.23    & Blends   \\
2753-54234-150 & 230.47445    & 12.39628     & 67.63    & Blends   & 1789-54259-351 & 190.44328    & 10.21984     & 22.16    & Bad Spectrum \\
346-51693-429  & 243.85497    & 0.65564      & 52.18    & Blends   & 348-51696-444  & 248.53931    & 1.00628      & 22.06    & Blends   \\
1383-53116-293 & 221.50511    & 35.31970     & 51.83    & Blends   & 1010-52649-301 & 159.86438    & 52.47833     & 21.98    & Blends   \\
1578-53496-55  & 241.85655    & 28.03230     & 51.75    & Blends   & 1740-53050-182 & 140.44511    & 10.74848     & 21.95    & Blends   \\
2607-54184-354 & 203.65768    & 16.62918     & 51.57    & Blends   & 653-52145-313  & 4.71200      & -10.37686    & 21.75    & Velocity Structure \\
2141-53764-533 & 222.01011    & 29.91320     & 47.87    & Blends   & 1013-52707-127 & 170.24723    & 53.19049     & 21.49    & Bright HII Region \\
2208-53880-85  & 246.49823    & 14.44520     & 46.48    & Velocity Structure & 1998-53433-304 & 159.05510    & 37.32432     & 21.18    & Wolf–Rayet \\
755-52235-410  & 115.89659    & 32.17637     & 44.88    & Blends   & 2753-54234-408 & 230.15202    & 13.56871     & 21.16    & Blends   \\
2370-53764-120 & 148.09108    & 18.06227     & 42.26    & Blends   & 541-51959-230  & 113.53775    & 31.97459     & 21.03    & Sky Contamination \\
1574-53476-78  & 246.57670    & 24.55564     & 40.27    & Blends   & 2885-54497-174 & 163.09376    & 8.98783      & 20.99    & Bad Spectrum \\
1716-53827-156 & 225.08889    & 11.55063     & 39.82    & Blends   & 2880-54509-269 & 185.47488    & 4.46184      & 20.98    & NI-5200  \\
489-51930-33   & 161.82787    & 65.11410     & 37.18    & Blends   & 2245-54208-263 & 202.76555    & 25.84251     & 20.90    & Strong H$\delta$  \\
1980-53433-27  & 169.71192    & 38.24940     & 35.84    & Blends   & 1876-54464-276 & 133.30146    & 63.49627     & 20.72    & Blends   \\
1708-53503-620 & 217.62451    & 13.65335     & 34.51    & Teacup AGN & 2208-53880-42  & 246.56421    & 14.24435     & 20.62    & Plate 2208 \\
1296-52962-221 & 124.98270    & 6.39953      & 31.92    & Blends   & 2791-54556-390 & 223.44661    & 20.27430     & 20.60    & Velocity Structure \\
348-51671-460  & 248.53931    & 1.00628      & 31.87    & Star     & 2208-53880-127 & 245.99632    & 14.62167     & 20.41    & Plate 2208 \\
891-52584-89   & 120.29245    & 32.75213     & 29.97    & Blends   & 1879-54478-496 & 148.67673    & 69.06061     & 20.21    & NI-5200  \\
2795-54563-133 & 234.11637    & 16.21026     & 29.62    & Blends   & 627-52144-321  & 247.42806    & 47.29304     & 20.16    & Bad Spectrum \\
2208-53880-265 & 245.79470    & 15.78185     & 29.47    & Plate 2208 & 1442-53050-520 & 173.30839    & 47.04056     & 20.13    & Bright HII Region \\
2170-53875-149 & 238.67295    & 17.30955     & 29.26    & Blends   & 339-51692-306  & 195.06886    & -2.90912     & 20.06    & Strong H$\delta$  \\
1713-53827-7   & 223.43381    & 9.09825      & 28.63    & Sky Contamination & 2208-53880-15  & 247.24661    & 14.57195     & 20.04    & Blends   \\
2598-54232-567 & 187.67913    & 18.76584     & 28.51    & AGN+He4686 & 2145-54212-388 & 222.88978    & 26.76765     & 19.86    & Wolf–Rayet \\
1369-53089-530 & 181.55612    & 45.33371     & 28.49    & Strong H$\delta$  & 1708-53503-256 & 216.14604    & 12.37499     & 19.81    & Blends   \\
2522-54570-408 & 239.28112    & 14.00622     & 28.26    & Blends   & 1877-54464-240 & 138.94872    & 64.26537     & 19.75    & Blends   \\
1768-53442-243 & 187.46343    & 14.06649     & 28.09    & Blends   & 776-52319-633  & 176.27600    & 62.31015     & 19.70    & Strong H$\delta$  \\
434-51885-137  & 116.84624    & 41.54075     & 27.69    & Blends   & 1463-53063-523 & 202.48142    & 47.23385     & 19.60    & NI-5200  \\
2208-53880-82  & 246.45008    & 14.48492     & 27.56    & Velocity Structure & 593-52026-182  & 233.43788    & 2.71444      & 19.55    & Blends   \\
457-51901-58   & 44.43610     & -8.51291     & 27.13    & Star     & 1786-54450-372 & 136.14741    & 62.38961     & 19.53    & Blends   \\
2765-54535-484 & 226.65602    & 15.93717     & 26.98    & Blends   & 486-51910-231  & 143.96497    & 61.35320     & 19.51    & Dusty    \\
1200-52668-24  & 138.81367    & 40.91724     & 26.93    & Dusty    & 280-51612-172  & 170.54637    & -0.07120     & 19.24    & Sky Contamination \\
2570-54081-85  & 121.76227    & 7.49888      & 26.54    & Blends   & 280-51612-213  & 170.24713    & -0.38365     & 19.24    & Blends   \\
1798-53851-174 & 199.16677    & 9.57150      & 25.84    & Bad Spectrum & 307-51663-111  & 220.04478    & -0.29398     & 19.23    & NI-5200  \\
892-52378-188  & 122.23627    & 35.27962     & 25.35    & Blends   & 2207-53558-27  & 249.43996    & 16.59545     & 19.21    & Velocity Structure \\
1359-53002-195 & 155.93396    & 41.06743     & 24.89    & Blends   & 401-51788-423  & 25.06214     & 0.41179      & 19.13    & Blends   \\
1627-53473-611 & 188.57624    & 8.23982      & 24.34    & Wolf–Rayet & 1581-53149-470 & 235.10316    & 32.86590     & 19.03    & SNIa     \\
575-52319-200  & 155.91516    & 4.18628      & 24.34    & NI-5200  & 638-52081-281  & 316.36185    & -7.57458     & 19.02    & Blends   \\
931-52619-573  & 125.66402    & 31.50058     & 24.26    & Blends   & 1677-53148-350 & 226.10314    & 44.41580     & 18.89    & Velocity Structure \\
1608-53138-173 & 175.89681    & 11.95587     & 23.98    & Bad Spectrum & 2763-54507-360 & 220.84255    & 16.21258     & 18.79    & Strong H$\delta$  \\
1324-53088-240 & 210.72928    & 54.37413     & 23.90    & NI-5200  & 1418-53142-370 & 239.62236    & 35.22463     & 18.67    & AGN+He4686 \\
1946-53432-30  & 148.30424    & 30.85623     & 23.84    & Blends   & 551-51993-372  & 132.80788    & 50.67865     & 18.56    & Dusty    \\
617-52072-326  & 235.02012    & 57.25070     & 23.28    & Velocity Structure & 1740-53050-102 & 141.32669    & 10.50220     & 18.49    & Velocity Structure \\
2208-53880-124 & 246.05605    & 14.66926     & 23.17    & Plate 2208 & 277-51908-519  & 165.98395    & 0.37677      & 18.45    & BL Lacertae \\
1999-53503-224 & 183.90322    & 33.52926     & 22.87    & Blends   & 2430-53815-267 & 132.43307    & 12.29871     & 18.42    & SNIa     \\
1767-53436-367 & 185.73890    & 15.86261     & 22.82    & NI-5200  & 1269-52937-606 & 131.88764    & 31.02697     & 18.42    & Velocity Structure \\
410-51877-620  & 44.88340     & 0.85483      & 22.57    & Blends   & 2208-53880-281 & 245.33715    & 14.68445     & 18.36    & Plate 2208 \\
1721-53857-102 & 231.05243    & 8.54481      & 22.55    & Dusty    & 1753-53383-252 & 170.07089    & 13.58969     & 18.31    & Dusty    \\
1740-53050-315 & 139.70837    & 10.57356     & 22.39    & Blends   & 2477-54058-389 & 159.76004    & 23.51568     & 18.20    & Strong H$\delta$  \\
845-52381-246  & 187.91661    & 3.94210      & 22.29    & Bright HII Region & 2745-54231-4   & 214.19363    & 14.70976     & 18.17    & Low-z Starburst \\
1835-54563-29  & 233.26141    & 4.20341      & 22.28    & Velocity Structure & 750-52235-326  & 358.94087    & 15.87922     & 18.00    & Blends   \\
    \bottomrule
    \\
  \end{tabular}}
    \caption{ The positions and visual classification labels of the top 100 SDSS galaxy outliers ranked by their negative log probabilities $-\log p$. }
  \label{tab:top100}
\end{table*}

\newpage


\bibliography{main}{}
\bibliographystyle{aasjournal}



\end{document}